# Delete: Deep Lead Optimization Enveloped in Protein Pocket through Unified Deleting Strategies and a Structure-aware Network

—When you face some problems in drug discovery, just delete!

HaotianZhang[1*], Huifeng Zhao[1,3*], Xujun Zhang[1,2*], Qun Su[1], Hongyan Du[1], Chao Shen[1], Zhe Wang[1], Dan Li[1], Peichen Pan[1], Guangyong Chen[2], Yu Kang[1*], Chang-yu Hsieh[1*], Tingjun Hou[1*]

[1] Innovation Institute for Artificial Intelligence in Medicine of Zhejiang University, College of Pharmaceutical Sciences, Zhejiang University, Hangzhou 310058, Zhejiang, China
[2] Zhejiang Lab, Zhejiang University, Hangzhou 311121, Zhejiang, China
[3] Hangzhou Carbonsilicon AI Technology Co., Ltd, Hangzhou 310018, Zhejiang, China

**Corresponding authors**
  **Tingjun Hou**
  **E-mail:** tingjunhou@zju.edu.cn
  **Chang-Yu Hsieh**
  **E-mail:** kimhsieh@zju.edu.cn
  **Yu Kang**
  **E-mail:** yukang@zju.edu.cn

## Abstract

Drug discovery is a highly complicated process, and it is unfeasible to fully commit it to the recently developed molecular generation methods. Deep learning-based lead optimization takes expert knowledge as a starting point, learning from numerous historical cases about how to modify the structure for better drug-forming properties. However, compared with the more established *de novo* generation schemes, lead optimization is still an area that requires further exploration. Previously developed models are often limited to resolving one (or few) certain subtask(s) of lead optimization, and most of them can only generate the two-dimensional structures of molecules while disregarding the vital protein-ligand interactions based on the three-dimensional binding poses. To address these challenges, we present a novel tool for lead optimization, named **Delete** (**De**ep **le**ad op**t**imization **e**nveloped in protein pocket). Our model can handle all subtasks of lead optimization involving fragment growing, linking, and replacement through a unified deleting (masking) strategy, and is aware of the intricate pocket-ligand interactions through the geometric design of networks. Statistical evaluations and case studies conducted on individual subtasks demonstrate that Delete has a significant ability to produce molecules with superior binding affinities to protein targets and reasonable drug-likeness from given fragments or atoms. This feature may assist medicinal chemists in developing not only me-too/me-better products from existing drugs but also hit-to-lead for first-in-class drugs in a highly efficient manner.

# Introduction

Drug discovery has long been recognized as a time-consuming, costly and risky process[1]. Computer-aided drug design (CADD) has been regarded as a more rational design approach to discover, design and develop therapeutic chemical agents[2]. In general, drug development involves four stages: hit identification, lead optimization, drug candidate selection, and drug approval[3,4,5]. Although a hit may have low potency, lead optimization strategies, as illustrated in Figure 1A, which include linker design, fragment elaboration, scaffold hopping and side-chain decoration, can improve drug-forming properties to develop a candidate drug[6]. Lead optimization was usually achieved by different strategies, such as library search based on similarity comparison and free energy calculation in the traditional context, such as fragment molecular orbital[7] (FMO) for linker design, AutoGrow[8] for fragment elaboration, and LigandScout[9] for scaffold hopping.

Another important application of lead optimization is to develop me-too/me-betters, which are chemically distinct enough to obtain new patents but share the same or better potency than first-in-class drugs[10]. This approach allows companies to establish intellectual property and recover development costs without requiring target validation and posing less risk than first-in-class drug discovery[11]. In history, many me-too/me-betters exemplify the scenarios we discuss here by utilizing lead optimization strategy, such as the fifth marketed drug for cholesterol, atorvastatin[12], is superior to the first-in-class lovastatin; and an antibacterial drug, levofloxacin, is superior to the first-in-class norfloxacin[13]. Although the me-too/me-better development paradigm has been a favorite of the pharmaceutical industry, CADD methods for me-too/me-better drug design have not been widely applied, and the development of me-too/me-betters still relies on experienced chemists[14,15]. This can be attributed in part to the fact that traditional lead optimization methods are limited by the crude definition of similarities between compounds and the restricted chemical space expressed in pre-prepared compound libraries.

With the explosion of crystallized data and the advancement of artificial intelligence (AI), data-driven-based similarity comparison has been used for enhancing the traditional lead optimization approaches[16], but this promotion has not brought essential changes in applications. Alternatively, molecular generation, with the support of AI generated content (AIGC) technology, has revolutionized not only for *de novo* design but also for lead optimization. Currently, there are four representatives, DeLinker[17], DeepFrag[18], DeepHop[19], and GraphScaffold[20], have been developed for the corresponding lead optimization sub-tasks (i.e., linker design, fragment elaboration, scaffold hopping, and side-chain decoration), respectively. However, in real-world applications, suppose that all of these strategies are needed, chemists must re-train four task-specific models on their own curated active compound data, which is laborious and far from intelligent generation. Developing a universal framework for all sub-tasks seems to be a challenging but exciting direction.

Inspired by masked graph modelling[21] and pre-training strategies[22] for Graph data structure, we introduce a series of deleting strategies, three enhanced and four task-specific, for constructing a general lead optimization model, titled **Delete** (**De**ep **le**ad op**t**imization **e**nveloped in protein pocket). Moreover, most previous deep learning (DL)-based pioneers for lead optimization have not taken the detailed target protein information into consideration, resulting in a larger chemical space rather than an active-centric cluster. When applying these models in drug discovery, additional tools such as quantitative structure-activity relationship (QSAR) models and molecular docking are required to reduce the scope of potential compounds. To address this issue, geometric neural networks were applied to constrain the spatial and energetic features of pocket environments in the chemical space. Besides, the equivariance design embedded in the model enables reliable conformation generation, adding an additional dimension of prediction to molecular generation.

To our best knowledge, Delete is the first structure-based model that can be applied to all the lead optimization tasks through a convenient unified deleting strategy. Based on the geometric and topological design of neural networks, Delete can learn the physical interaction between pocket residues and ligands in a physicist manner, and predict the binding poses of generated conformations directly in the given pocket, thus relieving the need to dock the generated molecules in the pocket required by the previous methods. The comprehensive statistical evaluation of the generated compounds on each sub-task demonstrates that the molecules generated by Delete not only maintain the same drug-forming properties, but also possess promoted binding affinities, highlighting the plausibility of applying a structure-based scheme in the lead optimization problem. The individual case study on each task reproduces several classic examples in drug design, further boosting the credibility of the Delete model.

## Results and Discussion

Delete provides a one-stop solution for lead optimization, which fully considers the chemical environments and geometric profiles within protein pockets. To demonstrate its potential, we conducted a comprehensive analysis of optimized molecules in four subtasks, namely linker design, fragment elaboration, scaffold hopping, and side-chain decoration. In the machine learning (ML) community, the requirement is raised naturally for comparing the newly developed method to its predecessors to indicate how much improvement it makes. However, only under the linker design is there a model similar to our work setting, DiffLinker[23], which could generate the 3D conformations of linked molecules inside a pocket. For the other three tasks, Delete is the first model of the 3D structure-aware kind, so we focus on the real-world lead optimization cases to compensate for this discrepancy and showcase for medicinal chemists of interest about how to employ Delete to aid their experimental intuition, accelerating the fast-follow and me-too development, a problem that is of vital importance but has been overshadowed by *de novo* drug design for a long time.

Two datasets were used for assessing Delete: one is CrossDock[24], which is enriched by the docking method and is a conventional choice for developing 3D pocket-aware DL models[25,26]; the other is BindingMoad[27], which contains only X-ray crystallized conformations that are closer to real-world scenarios. Three versions of Delete were trained both on these two datasets: Delete-C was trained on CrossDock, Delete-M was trained on BindingMoad, and Delete-CM was finetuned on BindingMoad after being pre-trained on CrossDock.

**Linker Design**

The purpose of linker design is to bridge two lead fragments through an intricate linker structure, resulting in improved binding affinities of the entire molecule via the famous phenomenon of superadditivity[7]. This optimization task is challenging due to the need to maintain the binding patterns of fragments to key residues in the linked molecules, which could be alleviated by the direct design of the linker inside a protein pocket[28]. Currently, DiffLinker is the only available model of this kind. **Table 1** shows the binding affinities and drug-forming properties of the generated molecules from three versions of Delete and DiffLinker. Notably, two affinity metrics

(Docking Affinity and Scoring Affinity) calculated by AutoDock-Vina are presented[29]. Docking Affinity is obtained based on the docking conformations, which reflects the true binding affinity more closely; while Scoring Affinity is calculated based on the directly generated conformations from DL models, which demonstrates the capability of these models to capture protein-ligand binding geometries. On the BindingMoad set, the ranking in terms of binding affinity is Delete-CM > Delete-M > BindingMoad > DiffLinker, indicating that DiffLinker cannot always optimize the molecule from lead fragments that can bind more tightly than its natural reference. The success rate of the molecules generated by Delete is higher than 90%, demonstrating that Delete is a powerful tool in linker design. Other drug-forming properties, such as QED, SA, Rule of Five (Lipinski) and LogP, are non-trivial prediction problems in computational chemistry[30,31]. Hence, we argue that these drug-forming properties of the generated molecules only need to fall within reasonable intervals, which Delete satisfies.

To further illustrate the potential of Delete in real-world drug discovery, we present a practical example from fragment-based drug discovery campaigns in the context of tuberculosis. Targeting inosine 5'-monophosphate dehydrogenase (IMPDH) presents a compelling strategy in drug design given its essential role as a key enzyme in the *de novo* biosynthesis of guanine nucleotides. Trapero et al.[32] have successfully discovered potent inhibitors for IMPDH through a comprehensive approach of fragment-based screening and structure-based design. During the fragment library screening, the authors identified a phenylimidazole derivative displaying low affinity towards IMPDH. To gain insight into the binding mechanism of this compound, X-ray diffraction analysis was performed, revealing that two molecules of the phenylimidazole derivative were concurrently engaged with the NAD pocket of IMPDH at a close molecular distance (**Figure 2A**). Through linking the two phenylimidazole molecules, the researchers succeeded in generating a novel compound (**Figure 2B**) that exhibits a remarkable improvement of over 1000-fold in IMPDH affinity compared with the initial fragment hit. To follow their footsteps, we sought to simulate the real design process of the most potent compound by deleting its linker atoms based on the structure of the initial fragment hit and employing Delete to generate the linker segment of the molecule (**Figure 2C-D**).

Remarkably, Delete successfully hit the origin compound by only generating a total of 76 different compounds. As shown in **Table 2**,

despite an average 2D similarity of only 0.5010 when compared with the original crystal compound, the generated compounds display a high 3D shape similarity of 0.9285, indicating that Delete can preserve the key geometrical features of the original compound while producing diverse molecules. The average SA of the generated compounds is notably high at 0.717, indicating the rationality of the generated compounds. When comparing the generated molecules with the original compound using Vina scoring, 65.8% of the generated molecules exhibit better scores (**Figure 2F**). The results of Figure 2G show that 60.5% of the generated samples gain higher estimated energy after conformational search, i.e., docking. **Figure 2E** exhibits the overlays of the four generated compounds and the corresponding 2D structures of the linker fragment. It demonstrates that the length of the linker generated by Delete is consistent with that of the linker in the original structure, and the generated compounds can accurately reconstruct the 3D structure of the original compound. For instance, as shown in the first example, the generated structure of the original compound is almost identical to that suggested by the crystal structure (RMSD). Furthermore, the other three examples indicate that Delete can provide more linkers with superior affinity and proper drug-likeness properties. To sum up, Delete possesses the capability to target reasonable and effective linkers with excellent efficiency, and generate molecules with improved scores, which will significantly expedite the drug linker design process.

**PROTAC Design**

As a specific task of linker design, proteolysis-targeting chimeras (PROTAC) design has garnered widespread interest. Recent research has explored the use of DL to accelerate PTOTAC design[33], but this approach primarily explores the 2D chemical space rather than rational design based on the interactions induced by protein pockets. So here, we employed the linker-trained version of Delete to design a series of PROTACs targeting the famous SMARCA2 case.

SMARCA2 is an ATPase subunit of the BAF (SWI/SNF) chromatin remodeling complexes, which controls genes involved in various biological processes such as DNA damage repair, DNA replication, and cell growth, division, and maturation[34]. Farnaby et al.[35] developed PROTAC degraders of SMARCA2 by employing a bromodomain ligand and recruitment of von Hippel–Lindau (VHL) E3 ubiquitin ligase. By isothermal titration calorimetry (ITC) experiments, the authors found that the binding of SMARCA2 and VHL by a poly(ethylene glycol)-based linker (PROTAC1) displayed 4.8 fold greater affinity. In order to not only mimic the binding conformation of PROTAC1 but also exhibit superior molecular recognition, they designed a linker by incorporating a benzylic group to form a π-stacking with the receptor Y98 residue (PROTAC2). The 2D structures and the overlay of two PROTACs are shown in **Figure 3A** and **Figure 3B.**

In order to evaluate the ability of Delete for the challenging task of PROTAC linker design, we excluded the known linker atoms of the PROTACs and retained solely the VHL and SMARCA2 binding fragments as the input for the Delete model (**Figure 3C**). Through this approach, we aim to investigate whether Delete could effectively generate promising linker structures in this such complex and demanding design task.

We directly generated 106 linker structures using Delete without any further optimization and filtering. After scoring these structures with AutoDock-Vina and comparing them with PROTAC2, we observed that 88.7% of the generated compounds had better scores (**Figure 3E**). The docking results surprisingly show that 98.1% of the compounds have better scores (**Figure 3F**). Furthermore, compared with the crystal structure of PROTAC2, the average 3D shape similarity of the generated molecules is as high as 0.8931, and their average SA is 0.5446, demonstrating that Delete can generate drug-like molecules with reasonable structures and higher activities for the PROTAC linker design task. Here, we present three generated PROTAC compounds that can well mimic the length and binding conformation of the linker in the crystal structure of PROTAC2 (**Figure 3D**). Interaction pattern analysis shows that these compounds demonstrate the ability to maintain the crucial CH−π interaction with Y98 observed in the crystal structure of PROTAC2. Summarily, Delete possesses the capability to PROTAC linker design, and the diversity of the generated compounds offer promising avenues for the design of novel linkers.

**Scaffold Hopping**

Scaffold hopping, a famous lead optimization strategy and also a well-known approach to develop me-too drugs, was first proposed by Gisbert[36] in 1999. Since then, a variety of computational methods have been developed[37-39] to spread this strategy. Most traditional methods rely on similarity comparison, such as bioisostere substitution or pharmacophore search[40]. In this study, we evaluated the performance of Delete on scaffold hopping under the Bemis-Murcko scaffold definition[41]. The results shown in **Table 1** indicate that not only the docking energies of the generated molecules are lower than those generated by CrossDock and BindingMoad but also the scoring energies are also lower, demonstrating the powerful generative capability of Delete to build favorable scaffolds inside the protein pockets. It is interesting to find that most of the drug-forming properties have a slight improvement in this task, though the BM scaffold removes the largest part of molecules as illustrated in **Figure 4D**. It seems that even only based several atoms scattered inside the pocket, Delete could still suggest potential structures for chemists. The impressive results of Delete on the most challenging scaffold hopping scenario implies that it would work well in the cases that removing fewer part of molecule, which is easier but closer to real-world applications.

We investigated the performance of Delete on scaffold hopping for kinesin Eg5, which is a crucial target for cancer chemotherapy drug development[42]. Ulaganathan et al.[42] identified and characterized an allosteric pocket of Eg5 along with an inhibitor BI8 with a nanomolar $K_d$, using X-ray methods (**Figure 4A-C**). To accomplish the scaffold hopping task, we employed the Bemis-Murcko scheme to select the core scaffold with its corresponding side chains and then deleted the scaffold of the BI8 compound, retaining only some tiny fragments (TFs) for generation (**Figure 4D**). The objective of this study is to investigate whether, using only TFs, Delete can construct reasonable scaffolds, generate compounds resembling the original structure, and create compounds that retain the geometrical similarity but with a novel skeleton.

We successfully generated a total of 131 structures, and surprisingly, our results demonstrate that Delete can produce compounds that closely resemble the original compound. The average 2D similarity of the generated structures is only 0.4700, implying that the generated scaffolds have changed to different structures compared to the original compound (previous work take 0.6 as the threshold[19]), while the 3D shape similarity was as high as 0.7878, indicating that the generated molecules retained the spatial characteristics of the original compound. By directly scoring the generated compounds using AutoDock-Vina and comparing them with the original compound, we found that the computed biological activity of the generated compounds was equivalent to or even higher than that of the original compound (**Figure 4F**). Additionally, we re-docked the generated and original compounds and compared the docking scores. The results show that 40.5% of the generated compounds achieve better scores than the originals (**Figure 4G**). **Figure 4E** exhibits the overlays of the four generated compounds, the interactions analysis and the corresponding 2D structures of the linker fragments. The first example generates a skeleton structure that closely resembles the original structure, with only one fluorine atom difference, and exhibits a near-perfect 3D conformational overlay with the original structure and maintains nearly all interactions. The three other examples illustrate three molecules generated by Delete, each of which possesses entirely different scaffolds, yet retains comparable 3D conformational characteristics with the original structure, maintaining most of the interactions observed in the crystal structure of the original compound and exploring even more interactions, such as the hydrophobic bonds with W127, L160, L214, F239 and the hydrophilic bonds with G217.

**Side-chain Decoration**

Side-chain decoration is another important lead optimization perspective, in which the foundation lies on growing side chains on privileged scaffolds[43] to explore possible interactions with residues while keeping the scaffold unchanged. Generating a series of scaffold-constraint compounds reduces the difficulty of synthesis and could retain the dominant structures of molecules that contribute to biological activity. **Table 1** presents the statistical results of the side-chain decoration under the BM scaffold setting, where the binding affinities and drug-forming properties are all advanced. The reason why QED and Rule of 5 achieve large improvement (14% and 25%) may contribute to the ease of decoration on the well-defined scaffold. To assess Delete under a more challenging circumstance, we selected a real-world case with a smaller scaffold.

D2 dopamine receptor (DRD2), a dopamine receptor subtype, plays a crucial role in the treatment of a variety of disorders, including Parkinson's disease, nausea, vomiting, and psychiatric disorders. However, many drugs targeting DRD2 often result in life-threatening side effects due to the interactions with other targets. Therefore, designing drugs based on the structure of DRD2 is a promising approach for developing psychiatric drugs. Recently, Wang et al.[44] solved the inactive conformational crystal structure of DRD2 in complex with the atypical psychiatric drug risperidone (**Figure 5A**), which revealed the presence of two sub-pockets positioned above and below the orthosteric site of the DRD2 receptor (**Figure 5B**). Specifically, these sub-pockets are respectively bound by the tetrahydropyridopyrimidinone and benzisoxazole moieties of risperidone. Furthermore, several significant residues such as W100, F110, W386, F390, Y408 and T412 were identified within this binding site (**Figure 5C**).

In this study, we utilized the piperidine structure of risperidone as a starting fragment for side-chain decoration to design novel drugs targeting DRD2 (**Figure 5D**). Through this approach, we generated 110 novel structures with a high degree of conformational similarity and molecular diversity compared to the X-ray crystallized compound. The average 2D and 3D similarities of the generated molecules are 0.4582 and 0.7422, respectively. Furthermore, the average SA and QED values of the generated molecules are 0.7215 and 0.6531, respectively, indicating the high synthesizable and druggable potentials of the generated compounds. Using AutoDock-Vina to directly score the generated structures, 23.6% of the molecules outperformed the original compound (**Figure 5F**). Furthermore, after docking and comparing the Top1 docking scores, 43.6% of the compounds exhibit higher binding capability (**Figure 5G**), suggesting the efficacy of our side-chains decoration approach for designing drugs targeting DRD2. **Figure 5E** exhibits the overlays of the four generated compounds and the origin compound. Despite the distinct 2D structures, the generated side chains from all four examples well reconstruct the 3D conformations of the side chains in the original compound. The interaction pattern analysis results demonstrate that all four structures maintain the interaction with W100, an important interaction with the inactive DRD2 receptor mentioned by Wang et al.[44], as well as most of the other interactions. Additionally, the generated structures enable better fitting within the pockets and thereby exhibit more interaction bonds, such as the hydrophobic bonds with L94, F198 and F389 and the hydrophilic bonds with D114 and S197.

**Fragment elaboration**

Fragment elaboration and side-chain decoration are two methods used in lead optimization to improve the pharmacological properties of drug candidates. While there is some overlap between the two techniques, fragment elaboration focuses on expanding a larger functional group that contains more pharmacophores to fill the empty sub-pockets, whereas side-chain decoration involves the optimization of 4-5 sites of the given scaffold. The results of fragment elaboration in **Table 1** keep similar to the other three tasks, the binding affinities are advanced and the drug-forming properties locate in a slight variance interval compared with the test set. In this task, recovery of lead optimization on the β1-adrenergic receptor (Adrb1) is the real-world case to demonstrate the potential of Delete in fragment elaboration.

Adrb1 is a classic drug target with a well-established structure that has been extensively studied over the years[45]. Adrb1 antagonists are frequently used in cardiovascular medicine, as well as in other therapeutic areas, including migraine and anxiety[46]. Notable examples of these antagonists include cyanopindolol and carazolol (**Figure 6A**), which possess comparable ethanolamine backbones and distinct aromatic or heteroaromatic moieties. To investigate whether Delete could successfully reproduce the structure of cyanopindolol and whether it could further hit the heteroaromatic skeleton of other Adrb1 inhibitors, we started with the crystal structure of cyanopindolol (**Figure 6B**) and removed the heteroaromatic moiety, retaining solely the ethanolamine backbone as the foundation for fragment elaboration (**Figure 6C**).

We generated a total of 117 structures and successfully obtained the exact structure of cyanopindolol while even exploring the structure of carazolol, another inhibitor of Adrb1. It is shown that the molecules generated by Delete not only gain a high 3D conformational similarity to the original compound but also maintain a high diversity, according to the average 2D similarity of the generated molecules being 0.3471 and the 3D shape similarity being 0.8138. Furthermore, the average SA and QED of the generated molecules are 0.7118 and 0.7669, respectively, demonstrating their high synthesizable and druggable potential. By directly scoring the generated compounds using AutoDock-Vina and comparing them with the original compound, we find that the activity of the generated compounds is equivalent to or even higher than that of the original compound (**Figure 6F**). Additionally, we re-docked the generated and original compounds to compare the docking scores. The results show that 66.7% of the generated compounds achieve better binding capability than the originals (**Figure 6G**), which further provides the effectiveness of the generated compounds. As shown in **Figure 6B**, the heteroaromatic moiety of cyanopindolol predominantly engages in hydrophobic and hydrophilic interactions with V122, S211, F307 and N310 of Adrb1. Four of the generated compounds are shown as examples in **Figure 6E**, among which the first example completely replicates the structure of the original compound, exhibiting excellent 3D structural superposition with the original compound and accurately reproducing all interaction features. The second example successfully hits another strong and potent Adrb1 antagonist compound, carazolol, and gains more interactions than the original structure as well as a higher Vina score. The wet experiment also reported that the activity of carazolol was about times higher than that of cyanopindolol (0.03nM vs. 0.26nM). Notably, these two representative samples are not present in the training set. The other two examples show more novel and reliable possibilities of Adrb1 inhibitors.

**Analysis of Generated Conformations**

Since Delete possesses the 3D generation capability, its generation of molecules is accompanied by the prediction of the binding conformation within the pocket. Previous experiments have focused on testing the chemical properties of the generated molecules, but this experiment emphasizes the plausibility of the geometries of the generated samples. Since it is impossible to get all the X-ray crystallized complex structures for our generated molecules, a feasible approach is to obtain the near-natural conformations by the docking method. Molecular docking was performed on the generated compounds and original ligands, and two RMSDs were computed: the RMSD of the crystallized conformations with the re-docked conformations for the original ligands (BindingMoad dataset); the RMSD of the docked conformations with the directly generated conformations for the model-generated molecules. The results shown in **Table 3** demonstrate that the two types of RMSD are quite similar, with the mean values within 2Å, which is a typical threshold for successful docking in virtual screening[47]. The generation of the conformations similar to the crystal structure confirms that Delete has learned physically meaningful protein-ligand interactions from a large amount of structure data to match the atoms to the appropriate potential low within the pocket.

# Limitations

Each model has its own limitations, which often rely on the underlying assumptions made in designing the model. All models are wrong, but some are useful, coined by the statistician, George Box[48]. In Delete, the basic assumption is the rigid generation assumption, i.e., the addition of atoms to the lead compound does not affect its position. This rigid constraint, although somewhat seemingly strong, still has its plausibility. The molecule obtained by rigid addition ensures that each step of atomic expansion fills a corresponding part of the pocket, resulting in an improvement of binding affinities with targets. If the structures change with the molecular growing process, the atoms that are suitable at the beginning will leave their proper positions during the change, and it cannot be guaranteed that the completed molecule fits the pocket perfectly. The discussion about which model based on which assumption is better needs to be further explored by future researchers.

## Conclusions

Delete is an all-in-one solution for lead optimization in drug discovery, made possible by the introduction of the 3D molecular generation framework and the unified deleting strategies. This framework enables a single model to perform multiple tasks of lead optimization with excellent efficiency. Additionally, the embedding of geometric neural networks allows for the simultaneous prediction of near-natural conformations of complete molecules, unifying molecular generation and conformation generation. Comprehensive in-silico experiments conducted under four different lead optimization scenarios have confirmed the capabilities of Delete. It can be employed not only to optimize lead compounds for first-in-class drugs but also to assist medicinal chemists in developing me-too/me-better products by structural modification/substitution from existing drugs. It is expected to see that Delete would be experimentally verified through feature real-world drug discovery campaigns.

## Method

### Dataset Construction

Delete has been trained on two datasets, one is CrossDock[24], and the other is BindingMoad[27]. CrossDock is a recognized benchmark for existing pocket-aware 3D *de novo* design models[26,49,50], which is curated from molecular cross docking results. Although docked conformations do not share exactly the same patterns with X-ray crystallized structures in principle, it is acceptable to enrich the dataset using physical tools since crystal data is far from saturated. In contrast, the

BindingMoad dataset, which is the evaluation choice for DiffLinker[23], consists of all crystal structures. After data processing, the CrossDock dataset has 100,000 training pocket-ligand pairs, while the BindingMoad has 35,516 training pairs.

The pocket-ligand pairs are needed to be further processed to become training data for lead optimization. We comprehensively investigate the previous lead optimization methods and select the corresponding representative approach to obtain sub-task data of lead optimization. In the linker design task, matched molecular pair analysis[51] (MMP) is performed to cut acyclic single bonds twice; in the fragment elaboration, functional fragments are obtained from one cut of acyclic single bonds; in the scaffold hopping, the scaffold is derived as Bemis-Murcko scaffold[41]; in the side-chain decoration, the side chains are all terminal acyclic groups. Besides, we also provide some other decomposition methods in the program, such as using the BRICS[52] rule to obtain fragments and using ScaffoldNetwork[53] to obtain intermediate scaffolds.

**Unified Deleting Framework**

Self-supervised learning has made significant progress in deep learning, substantially boosting the model's performance on many tasks[54]. As a part of self-supervised learning, the masking strategy aims to mask and recover whole, parts of, or merely some features of its original input[55-57]. In natural language processing, researchers mask the context and recover the marked words or phrases in the pre-training phase, so the resulting model can be used either as a pre-training encoder or directly for text generation[58,59]. In computer vision, researchers randomly mask pixels in images and recover them with Encoder-Decoder architecture[55]. In summary, pre-trained with masking strategies, models can automatically learn the intrinsic structure in the data itself on large amounts of unlabeled staff, which in turn improves their performance on downstream tasks. In the context of molecular data pre-training, there have also been some developments in masking strategies. For instance, Zhou et al. mask-recover the atomic type[60], while Jiao et al. recover the noised atomic coordinates[61]. Similarly, masking strategies have been widely used in molecular generation. GraphAF[62] uses masking strategies to make the Flow model satisfy autoregressive properties and can be trained in parallel; C5T5[63] introduces IUPAC mask to achieve modification of the input compound; Mask Graph Modelling[21] randomly mask-reproduces atoms and bonds in the molecule, which enables gradual transition of seed molecule to the structurally novel compound.

Based on the inspirations outlined above, we propose a unified deleting (masking) strategy to create a single model capable of handling all subtasks (Figure 1C). Our strategy comprises seven deleting techniques, of which the first three are the random mask, spatial mask, and topological mask. These techniques are employed to enhance the training process. The remaining four masks, namely the linker mask, fragment mask, scaffold mask, and side-chain mask, correspond to four specific subtasks in lead optimization. $N$ represents the number of atoms with the molecule, and $\sim_R S$ denotes unit sampling from set $S$:

**Random**: a probability $p \in [0,1]$ is predefined, and then each atom is masked independently according to the drop-out probability $p$.

**Spatial**: a focal atom $f_i$ is randomly sampled, i.e., $f_i \sim_R [n_1, n_2 ..., n_N]$, and atoms within the predefined distance threshold are masked. This masker helps the model capture spatial relationships.

**Topological**: Started with a random atom $f_i$ and predefined discard number $m$, atoms $[f_i, ..., f_i + n_m]$ involved in the topological path determined by the breadth-first search (BFS) are masked. This masker helps the model capture molecular topological structure.

**Fragment**: The fragment is determined by cutting one of the acyclic single bonds that are not within functional groups. The smaller one from the resulting two fragments is chosen to be masked. This masker is designed for fragment elaboration/replacement.

**Linker**: The linker mask, derived from double cuts of acyclic single bonds that were not with functional groups, is defined as the fragment that is sandwiched between two resulting fragments. This masker is designed for linker design.

**Scaffold**: The scaffold masker definition follows the classical Bemis-Murcko[41] approach, which is described as masking structures except for all terminal acyclic side chains. This masker is designed for scaffold hopping.

**Side-chain**: The side-chain masker is defined as making all terminal acyclic side chains. This masker is designed for side-chain decoration.

To train a target-specific model, we combine three enhanced maskers with each task-specific masker. The resulting model is then evaluated under different task scenarios using its corresponding trained parameters.

The workflow of Delete is simplified and shown for understanding in **Figure 1.** The module composition of Delete is outlined as follows.

**Protein-Ligand Interaction Module**

To better extract interaction features, the protein pocket is characterized as a surface profile consisting of triangle patches[64]. Then the pocket graph $\mathcal{G}_p$ is built up, where nodes are vertices of the triangular surface and edges are determined by the nearest neighbors. The ligand is characterized as a chemical graph structure, denoted $\mathcal{G}_l$, where nodes are atoms and edges are covalent bonds. Firstly, these features with non-uniform dimensionality and physical intuition should be mapped to the same space. To achieve this, two embedding modules are utilized, which is outlined as follows:

$$\mathcal{G}_{pl}(n_i', \vec{n_i'}) = \text{NodeEmbed}(\mathcal{G}_l\left(n', \vec{n'}\right), \mathcal{G}_p(n', \vec{n'}))$$

$$\mathcal{G}_{pl}(e_i', \vec{e_i'}) = \text{EdgeEmbed}(\mathcal{G}_l\left(e', \vec{e'}\right), \mathcal{G}_p(, e', \vec{e'}))$$

where $n_i, \vec{n_i}$ and $e_i, \vec{e_i}$ are chemical and geometric features of nodes and edges, $\mathcal{G}_p$ and $\mathcal{G}_l$ are pocket graph and ligand graph. After the embedding operation, they are mapped to a higher dimensional space, denoted as $n_i', \vec{n_i'}, e_i', \vec{e_i'}$. To learn the effect of pocket residues on the distribution of atoms within the pocket, the geometric message passing framework[50,65,66] is applied on the pocket-ligand graph $\mathcal{G}_{pl}(n', \vec{n'}, e', \vec{e'})$.

$$\mathcal{G}_{pl}(m_i, \vec{m_i}) = \text{Message}(\mathcal{G}_{pl}(n', \vec{n'}, e', \vec{e'}))$$

$$\mathcal{G}_{pl}(h_i, \vec{h_i}) = \text{Update}(\mathcal{G}_{pl}\left(n', \vec{n'}, m_i, \vec{m_i}\right))$$

where $h_i, \vec{h_i}$ are extracted interaction features assigned on nodes. At this point, each node of $\mathcal{G}_{pl}$ fully perceives the influence of its geometric neighbors.

**Attachment Point Enumeration**

The first step in generating a new compound is to predict potential attachments on the existing lead compound. To do this, we enumerate all the atoms within the lead and predict their attachment scores, denoted as $p_{a_i}$. A high score indicates a higher possibility of attachment. The overview of attachment prediction goes as follows:

$$(a_i, \vec{a_i}) = f(W_{a1} \cdot h_i, \vec{h_i} \cdot W_{a2})$$

$$a'_i = \| \vec{a_i} \|_2 + f(a_i \cdot W_{a3})$$

$$p_{a_i} = \sigma(a'_i \cdot W_{a4})$$

where $W_{a1,2,3,4}$ are linker transform matrixes, $\sigma$ is the sigmoid function, $f$ is the active function, default is Leaky Relu.

**Atom Placement**

To place atom $i$ within pockets, the coordinate $\vec{r_i}$, atomic symbol $s_i$, and bonding relationship $b_{ij}$ with existing atoms $j$ should be generated. First, we generate the coordinate and relative position of atom $i$ using the following equations:

$$(x_i, \vec{x_i}) = f(W_{x1} \cdot h_i, \vec{h_i} \cdot W_{x2})$$

$$(w_i, \vec{w_i}) = f(W_{x3} h_i, \vec{h_i} W_{x4})$$

$$\vec{r_i} = r_{a_i} + \sum_{k=1}^{K} w_i^k \vec{x_i}^k$$

where $W_{x1}, W_{x2}, W_{x3}, W_{x4}$ are parameterized matrixes, $f$ is the active function, $w_i^k$ is the coefficient of k-th component of relative coordinates, $r_{a_i}$ is the coordinate of the attachment point. The coordinate of atom $i$ is determined by the coordinate of the attachment point and the relative position. Such a design reduces the error compared with the direct prediction of Cartesian coordinate[67], and avoids predicting 1-2, 13, 1-4 atoms in the local coordinate system[68]. Next, we predict the atomic symbol $s_i$ based on the interaction features of the pocket-ligand graph $\mathcal{G}_{pl}$ using the following equations:

$$(m'_i, \vec{m'}_i) = \sum_{k \in \text{kNN}(i)} \text{Message}(\mathcal{G}_{pl}(h_i, \vec{h_i}, e_i, \vec{e_i}))$$

$$(s_i, \vec{s}) = f(W_{s1} \cdot m'_i, \vec{m'}_i \cdot W_{s2})$$

where Message is the message passing module, $W_{s1}$ is the parameterized matrix. Finally, we predict the bonding relationship $b_{ij}$ with the existing atoms j using the following equations:

$$(o_{ij}, \vec{o_{ij}}) = f(W_{b1} \cdot [s_i || s_j || e_{ij}], [\vec{s_i}, \vec{s_j}, \vec{e_{ij}}] \cdot W_{b2})$$

$$(o'_{ij}, \vec{o'}_{ij}) = \text{TriAttention}(o_{ij}, \vec{o_{ij}})$$

$$(b_{ij}, \vec{b_{ij}}) = f(W_{b3} \cdot o'_{ij}, \vec{o'}_{ij} \cdot W_{b4})$$

Unlike previous work[68,69] that determines the bonding relationship with chemical rules (usually by OpenBabel), we predict it directly, which backward more information to the embedding module and reduces the potential error introduced by external bonding determination.

**Generation Complete**

There are two criteria for stopping generation, one is the given max number of atoms, and the other is automatically stop when the attachment probability of each atom is lower than a given threshold (e.g., 0.5).

**Other setups**

*Molecular Similarity.* The molecular similarity is the Tanimoto coefficient between two molecules, which is defined as follows:

$$\text{Tanimoto}(G_1, G_2) = \frac{|G_1 \cap G_2|}{|G_1 \cup G_2|} = \frac{|G_1 \cap G_2|}{|G_1| + |G_2| - |G_1 \cap G_2|}$$

where G is the molecular fingerprint, a common choice is topological fingerprints in RDKit[70].

*3D similarity.* 3D similarity metrics encompass shape similarity and electrostatic similarity[71]. Shape similarity is quantified using the shape Tanimoto distance[72], while electrostatic similarity is determined by the Carbo similarity[73] of the Coulomb potential overlap integral, which employs Gasteiger charges[74].

## Supporting Information

**Part S1**. Details of the design of the Delete modules.

## Acknowledgments

This work was financially supported by National Natural Science Foundation of China (22220102001), National Key Research and Development Program of China (2021YFF1201400),

## Tables

**Table 1.** The mean binding energies and drug-like properties[a] of the Top 5 molecules on the four lead optimization sub-tasks.

|  | CrossDock | Delete-C | BindingMoad | DiffLinker | Delete-M | Delete-CM |
|---|---|---|---|---|---|---|

|  | Linker Design | | | | | |
|---|---|---|---|---|---|---|
| Docking Energy (↓) | -8.888 | -10.235 | -8.540 | -7.992 | **-9.427** | **-9.454** |
| Scoring Energy (↓) | -7.786 | -8.765 | -6.590 | -5.972 | **-7.131** | **-7.208** |
| Hit Pocket (%) | | 96.44 | | 51.28 | **91.50** | **92.50** |
| QED (↑) | 0.434 | 0.438 | 0.493 | 0.477 | 0.459 | 0.459 |
| SA (↑) | 0.737 | 0.740 | 0.673 | 0.686 | 0.675 | 0.682 |
| Lipinski (↑) | 4.474 | 4.020 | 4.091 | 3.749 | 3.382 | 3.442 |
| LogP | 0.034 | 2.323 | 0.444 | 0.861 | 0.621 | 0.765 |
|  | Scaffold Hopping | | | | | |
| Docking Energy (↓) | -7.896 | -9.357 | -7.912 | | **-8.816** | **-8.871** |
| Scoring Energy (↓) | -6.583 | -7.533 | -4.591 | | **-5.202** | **-5.301** |
| Hit Pocket (%) | | 92.72 | | | **87.35** | **91.95** |
| QED (↑) | 0.434 | 0.459 | 0.337 | | 0.310 | 0.310 |
| SA (↑) | 0.675 | 0.689 | 0.583 | | 0.602 | 0.605 |
| Lipinski (↑) | 3.708 | 4.414 | 2.586 | | 2.844 | 2.894 |
| LogP | 0.465 | 0.554 | -0.724 | | -0.909 | -0.788 |
|  | Side-chain Decoration | | | | | |
| Docking Energy (↓) | -7.605 | -8.845 | -7.947 | | **-8.768** | **-9.118** |
| Scoring Energy (↓) | -6.389 | -7.216 | -5.490 | | **-6.436** | **-6.698** |
| Hit Pocket | | 91.17 | | | **77.68** | **86.77** |
| QED (↑) | 0.432 | 0.495 | 0.380 | | 0.438 | 0.432 |
| SA (↑) | 0.657 | 0.655 | 0.619 | | 0.622 | 0.616 |
| Lipinski (↑) | 3.842 | 4.248 | 2.766 | | 3.516 | 3.742 |
| LogP | 0.457 | 0.408 | -0.527 | | -0.347 | -0.495 |
|  | Fragment Elaboration | | | | | |
| Docking Energy (↓) | -8.781 | **-9.815** | -9.017 | | **-9.656** | **-9.889** |
| Scoring Energy (↓) | -7.720 | -8.333 | -7.758 | | **-8.320** | **-8.396** |
| Hit Pocket (%) | | 94.28 | | | **82.63** | **91.61** |
| QED (↑) | 0.444 | 0.406 | 0.557 | | 0.521 | 0.501 |

| | | | | | |
|---|---|---|---|---|---|
| SA (↑) | 0.663 | 0.645 | 0.671 | 0.669 | 0.665 |
| Lipinski (↑) | 4.276 | 3.689 | 4.347 | 3.936 | 3.905 |
| LogP | -0.326 | 1.043 | 0.529 | 1.857 | 1.654 |

[a]**Docking Energy and Scoring Energy** represent the binding energies of ligands to protein pockets (w and w/o binding conformation search); **Hit Pocket** represents the probability of that targets that have the generated molecules with higher binding affinity than their original ligands; **QED**[75] represents the **Q**uantitative **E**stimation of **D**rug-likeness, an integrative score to evaluate compounds' favorability to become a hit; **SA**[76] is the **S**ynthetic **A**ccessibility evaluation, a score to assess the ease of synthesis of compounds, a higher score corresponds to an easier synthesized molecule; **Lipinski**[77] is Lipinski's Rule-of-Five, a rule of thumb to evaluate drug-likeness, the higher the better; **LogP**[78] is the octanol-water partition coefficient, and logP values should generally be between -0.4 to 5.6, ideally between 1.35-1.8 for good oral intestinal absorption.

**Table 2.** The mean binding energies, drug-like properties and similarity analysis of five real-world cases[a].

| | Adrb1 | IMPDH | Smarca2 | Drd2 | Cox2 |
|---|---|---|---|---|---|
| Docking Energy (↓) | -9.090±0.758 | -9.004±0.628 | -14.304±0.620 | -11.660±0.922 | -12.003±0.575 |
| Scoring Energy (↓) | -7.275±0.937 | -7.770±0.573 | -13.893±0.626 | -9.970±1.064 | -11.206±0.928 |
| QED (↑) | 0.766±0.102 | 0.389±0.048 | 0.097±0.012 | 0.653±0.135 | 0.805±0.091 |
| SA (↑) | 0.711±0.080 | 0.717±0.043 | 0.544±0.020 | 0.721±0.072 | 0.834±0.046 |
| 2D. Sim. | 0.347±0.107 | 0.501±0.112 | 0.844±0.032 | 0.458±0.047 | 0.371±0.128 |
| Shape. Sim. | 0.813±0.040 | 0.928±0.018 | 0.893±0.006 | 0.742±0.034 | 0.889±0.021 |
| Esp. Sim. | 0.994±0.003 | 0.998±0.002 | 0.999±0.002 | 0.992±0.003 | 0.994±0.005 |

*a***Docking Energy, Scoring Energy, QED, and SA** are kept in the same definition as Table 1. **Shape Sim.** is Shape similarity and **Esp. Sim.** is electronic similarity.

**Table 3.** Docked RMSD analysis on the crystallized conformations (BindingMoad) and Delete-generated conformations.

|  | BindingMoad | Delete-C | Delete-M | Delete-CM |
|---|---|---|---|---|
| Linker-Design | 1.3877±0.6989 | 1.5886±0.7685 | 1.4504±0.6230 | 1.4695±0.6180 |
| Scaffold-Hopping | 1.6165±0.8265 | 1.5662±0.5952 | 1.7893±0.6326 | 1.7810±0.6328 |
| Fragment Growing | 1.3877±0.6989 | 1.444Å±0.6709 | 1.1444±0.5141 | 1.2621±0.5408 |
| Sidechains Decoration | 1.4971±0.8112 | 1.4681±0.5946 | 1.3912±0.6943 | 1.4902±0.7488 |

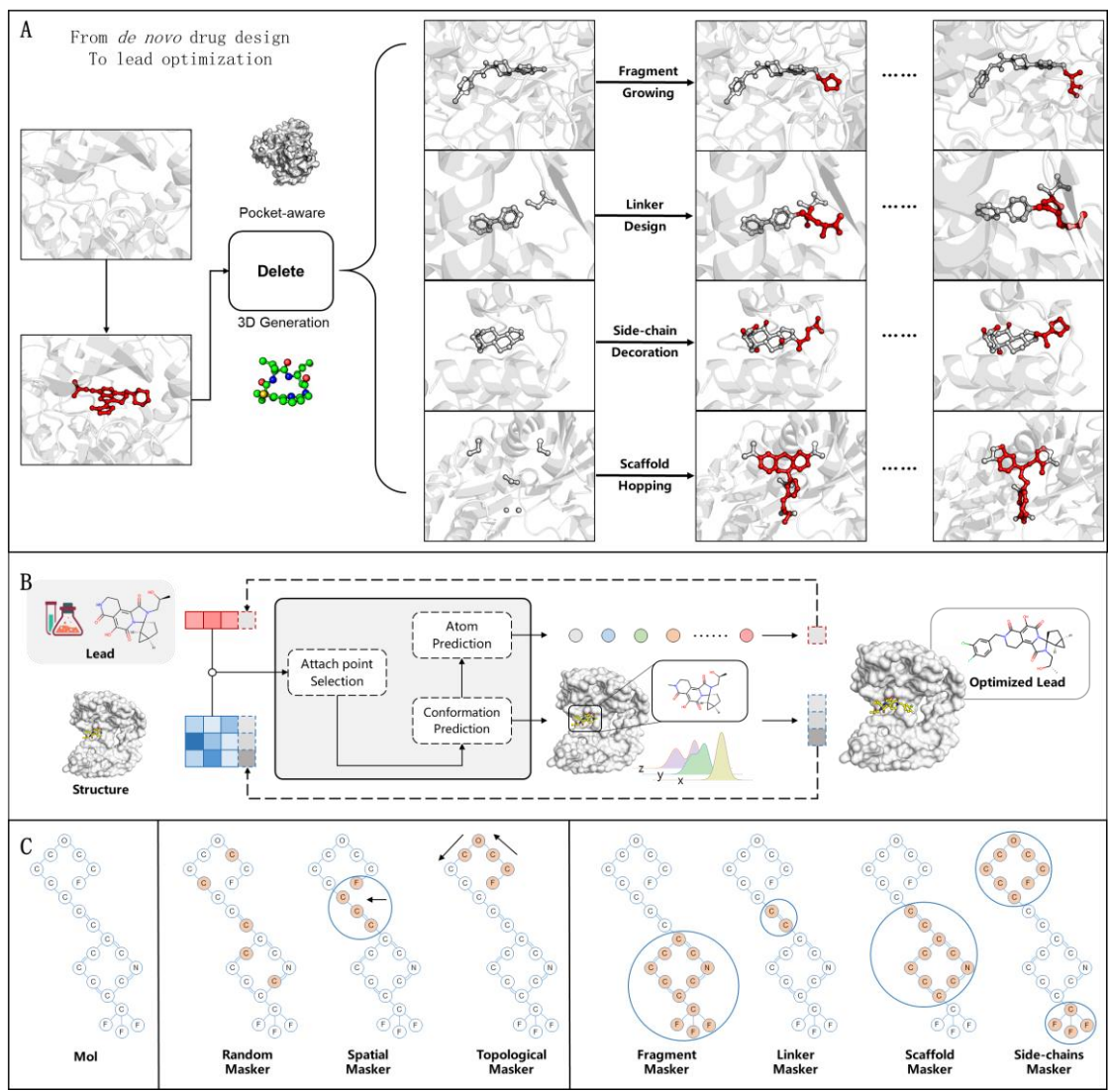

**Figure 1. Delete is a one-for-all solution for lead optimization.** A). Delete is a pocket-aware and 3D-generation model, which expands *de novo* design to four lead optimization tasks. B). The workflow of Delete. C). The unified deleting (masking) strategies illustrative, the first three are enhanced strategies while the last four are task-specific strategies.

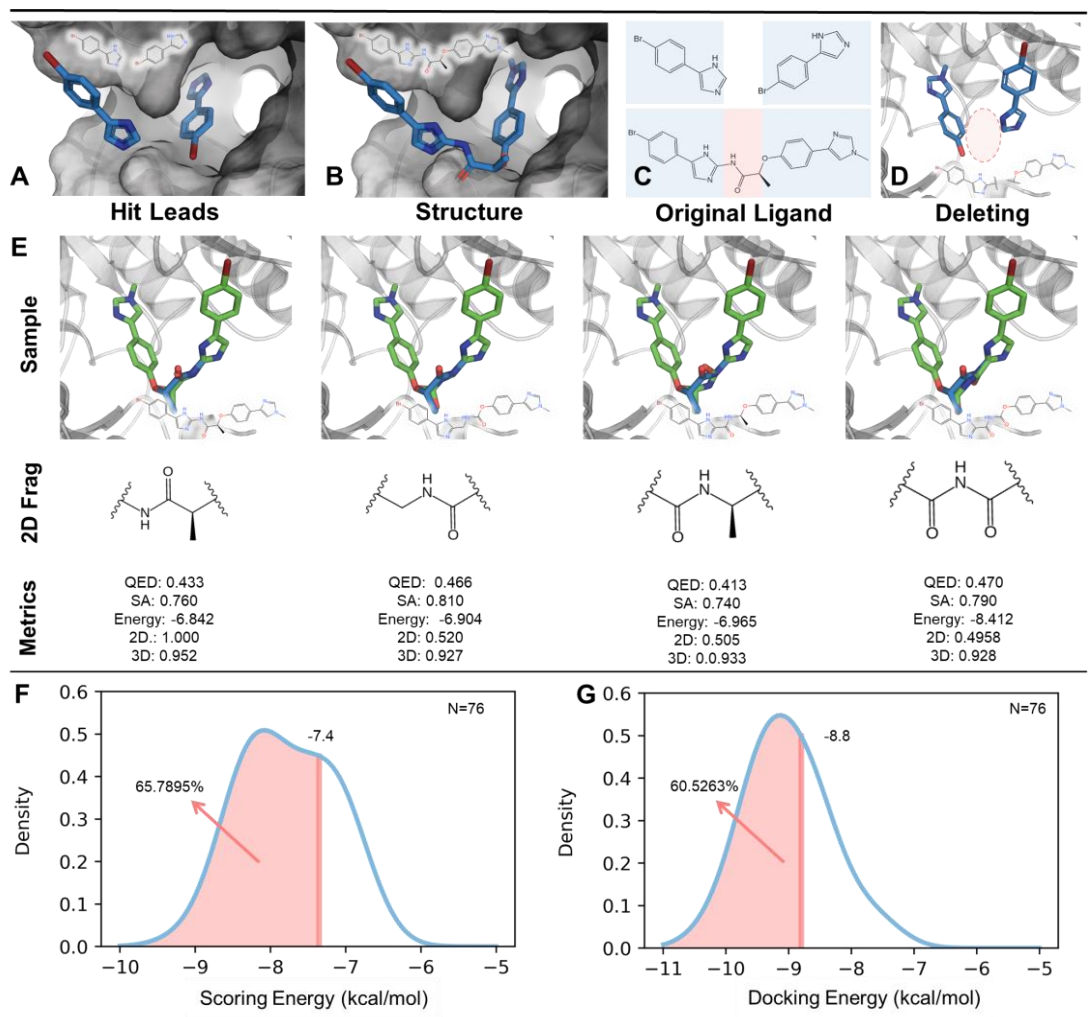

**Figure 2. Linker design case study: IMPDH.** A). Two phenylimidazole derivatives were concurrently engaged with the NAD pocket of IMPDH (PDB ID: 5OU2). B). Crystal structure of the original ligand (PDB ID: 5OU3). C). 2D chemical structure of phenylimidazole derivative and original compound. D). The starting point for Delete generation, where the red cycle is the potential growing space. E). Overlays of the four generated examples (green) and the original compound (blue), where metrics are molecular properties. F, G). The blue curve is the binding affinity distribution of generated compounds, where the orange line denotes the original compound. F is the docking energy, while G is the scoring energy.

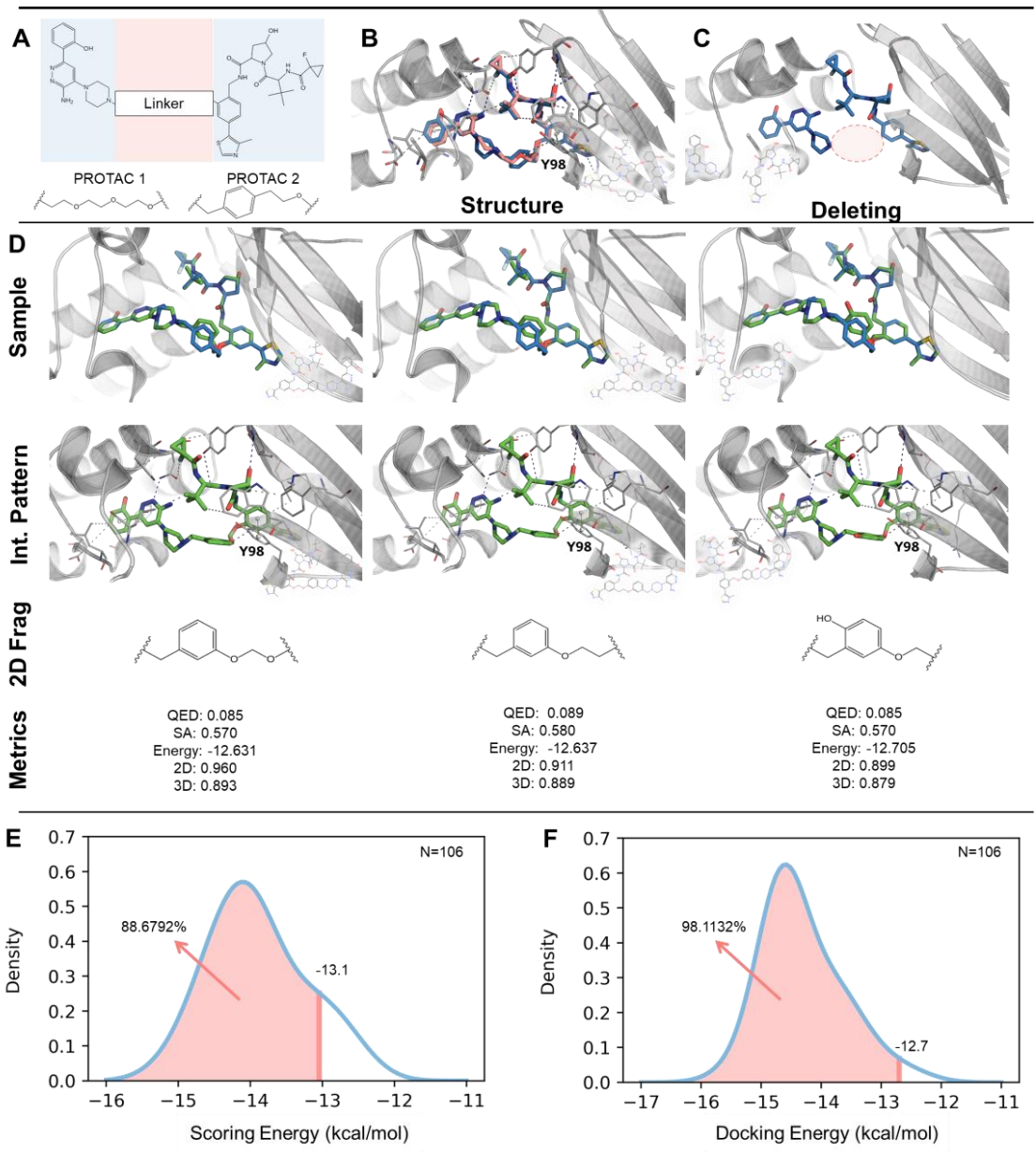

**Figure 3. PROTAC linker design case study: SMARCA2** A). 2D chemical structure of PROTAC 1 and PROTAC 2. B). Overlay of crystal structures of PROTAC 1 (PDB ID: 6HAY, red carbons) and PROTAC 2 (PDB ID: 6HAX, blue carbons). C). The starting point for Delete generation, where the red cycle is the potential growing space. D). Overlays of the three generated compounds (green) and the original compound (blue). The second line is the interaction pattern analysis of generated candidates, where important residues are identified with residue name-id. E, F). The blue curve is the binding affinity distribution of generated compounds, where the orange line denotes the original compound. E is the docking energy, while F is the scoring energy.

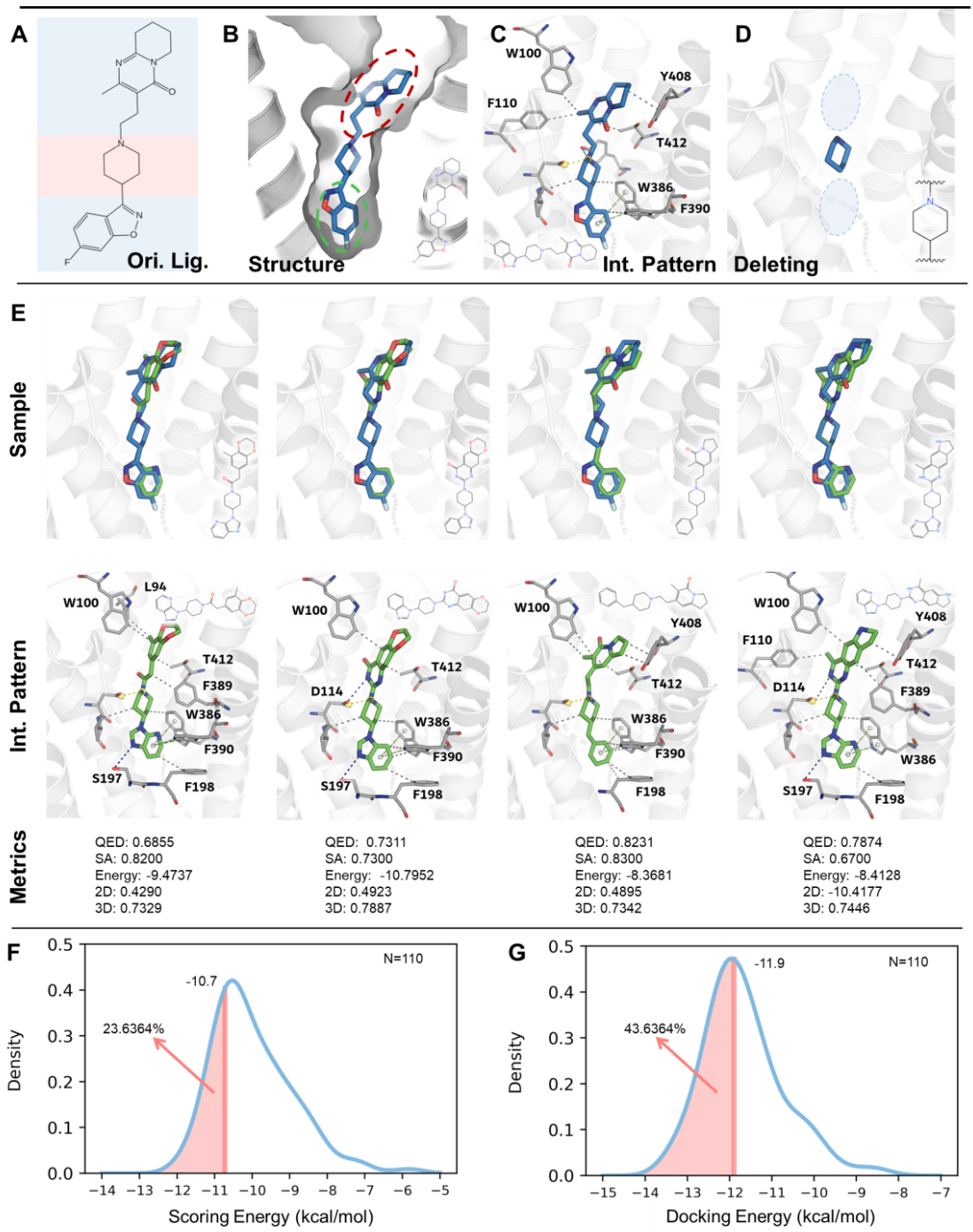

**Figure 4. Scaffold hopping case study: Eg5** A, B). 2D chemical structure of BI8 and its scaffold, the red represents the Bemis-Murcko backbones, and the blue part represents the TFs. C). Crystal structure of BI8 (PDB ID: 3ZCW). D). The starting point for Delete generation, where the red cycle is the potential growing space. E). Analysis of four generated molecules. F, G). The blue curve is the binding affinity distribution of generated compounds, where the orange line denotes the original compound. F is the docking energy, while G is the scoring energy.

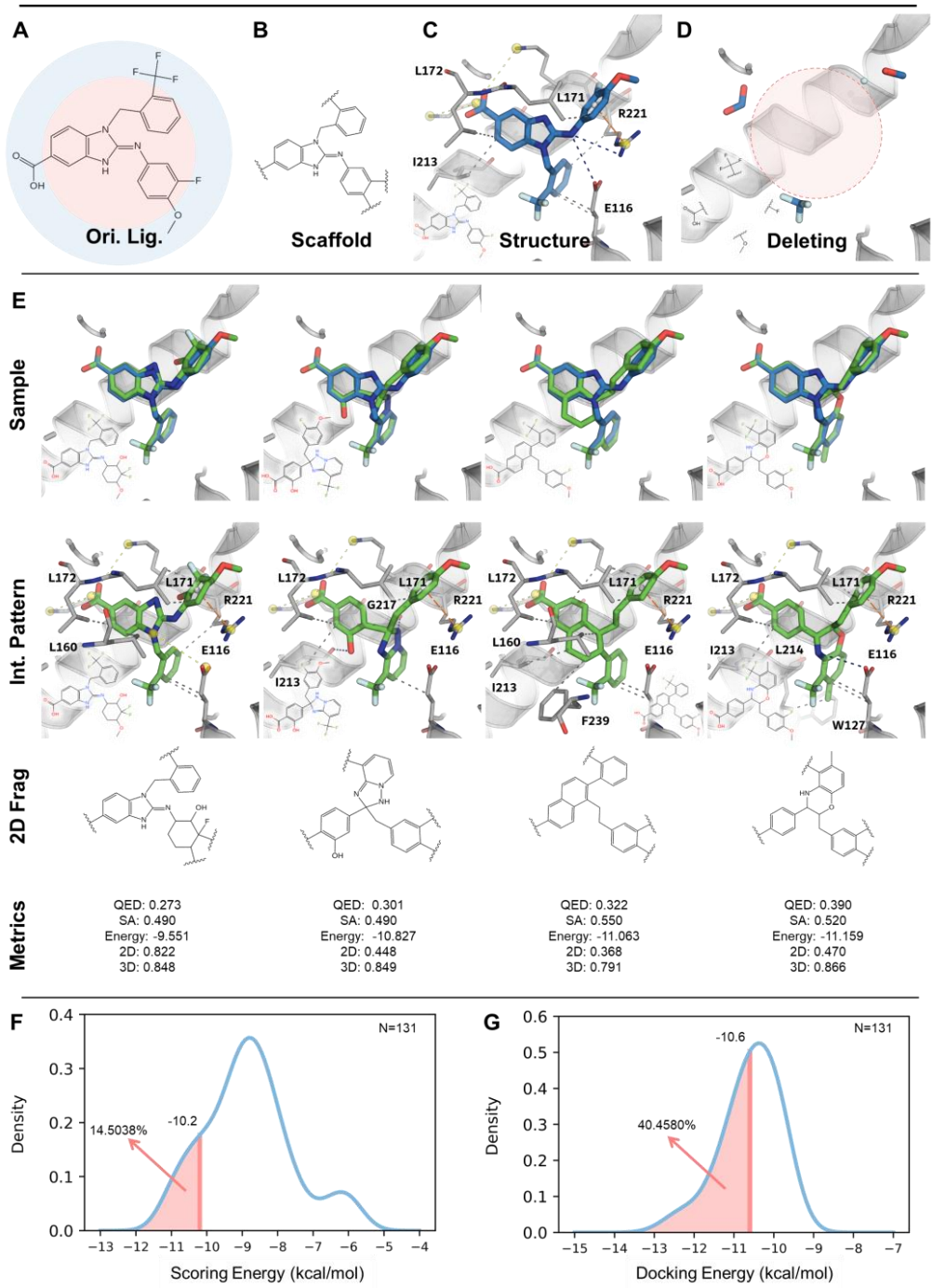

**Figure 5. Side-chains decoration case study: DRD2.** A). 2D chemical structure of risperidone. B). Crystal structure of DRD2's orthosteric site (PDB ID: 6CM4), along with risperidone, the red and green cycles locate two sub-pockets. C). Interaction analysis of risperidone. D). The starting point for Delete generation, where the red cycle is the potential growing space. E). Analysis of four generated molecules. F, G). The blue curve is the binding affinity distribution of generated compounds, where the orange line denotes the original compound. F is the docking energy, while G is the scoring energy.

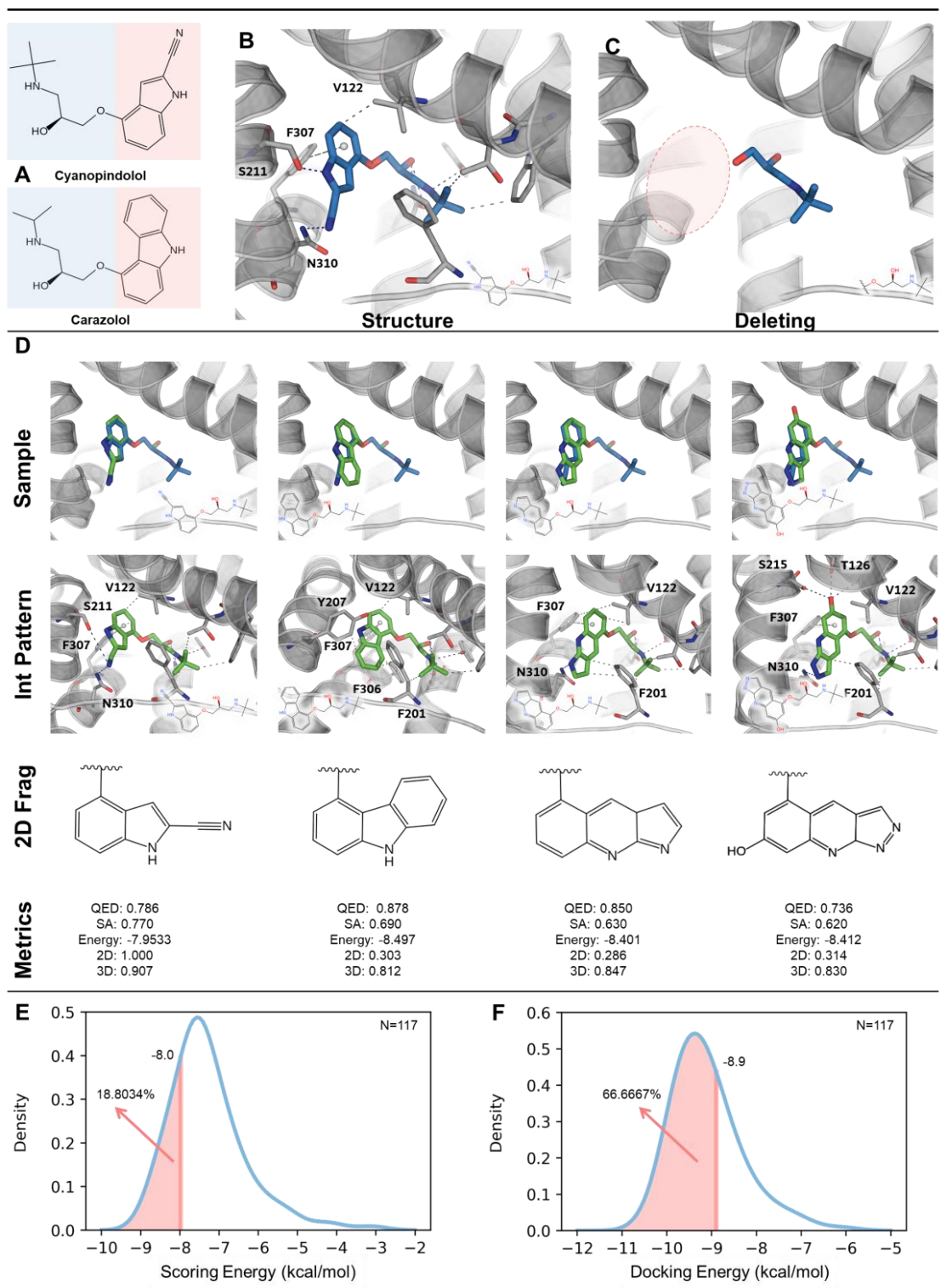

**Figure 6. Fragment growing case study: Adrb1.** A). 2D chemical structure of cyanopindolol and carazolol, the blue part represents the ethanolamine moiety, while the red part represents the heteroaromatic moiety B). Crystal structure of cyanopindolol (PDB ID: 2VT4). C). The starting point for Delete generation, where the red cycle is the potential growing space. D). Analysis of four generated molecules. E, F). The blue curve is the binding affinity distribution of generated

compounds, where the orange line denotes the original compound. E is the docking energy, while F is the scoring energy.